\documentclass[letterpaper,twocolumn,10pt]{article}
\usepackage{usenix-2020-09}

\usepackage{times}  
\usepackage{titlesec}
\usepackage{hyperref}
\usepackage{tikz}
\usepackage{amsmath, amssymb}
\usepackage{graphicx}
\usepackage{booktabs,tabularx,threeparttable,float}
\usepackage{xspace}
\usepackage{xcolor}
\usepackage[T1]{fontenc}
\usepackage{subfigure}
\usepackage{verbatim}
\usepackage{url}
\usepackage{enumitem}
\usepackage{listings}
\usepackage{multirow}
\usepackage[ruled,linesnumbered]{algorithm2e}
\usepackage{pifont}
\usepackage{mathtools}
\usepackage{fancyhdr}
\usepackage{caption}
\usepackage{xurl}
\usepackage{longtable}

\newcommand{\sysname}{\texttt{SiDP}\xspace}

\newcommand{\para}[1]{\smallskip\noindent\textbf{#1}}

\usepackage{filecontents}

\begin{document}

\date{}

\title{SiDP: Memory-Efficient Data Parallelism for Offline LLM Inference}

\author{
{\rm Alan Zhao}\\
scitix
\and
{\rm Cyril Y. He}\\
scitix
} 

\maketitle

\begin{abstract}
The rapid adoption of large language models (LLMs) has shifted a substantial portion of inference workloads into throughput-oriented offline regimes, where fully utilizing GPU compute requires large batch sizes. However, existing deployments face a structural tension. Data parallelism (DP) scales throughput well but replicates model weights, leaving limited GPU memory for key–value (KV) cache and constraining batch size. Model parallelism reduces per-device weights, but requires fine-grained synchronization that erodes DP’s independence and scheduling flexibility.
We present \emph{SiDP}, a memory-efficient data-parallel paradigm for offline LLM inference that treats weights as a bandwidth-backed shared resource inside a DP group. Instead of storing the full model on every GPU, \sysname organizes weights as a distributed pool: each layer is owned by a single GPU, and other replicas access its weights on demand via two complementary execution modes—a \emph{Weight-as-a-Service} (WaS) mode that streams remote weights over NVLink into a small cache in the large-batch regime, and a \emph{Compute-as-a-Service} (CaS) mode that ships activations to owners in the small-batch tail. 
Evaluated on NVIDIA H20, H200, and B200 GPUs with Qwen3-32B, Qwen2.5-72B, and Llama-3.1-70B, \sysname increases usable KV capacity by up to $1.8\times$ under the same configurations, and converts this into up to $1.5\times$ higher end-to-end throughput over baselines (vLLM) for offline workloads.
\end{abstract}

\section{Introduction}
\label{sec:intro}

The meteoric rise of large language models (LLMs)~\cite{ouyang2022gpt, liu2024deepseekv3, yang2025qwen3, llama-3.1, gemma_2024} has split inference workloads into two regimes. \emph{Online services}—such as chatbots~\cite{roller2021recipes}, coding assistants~\cite{chen2021evaluating}, and real-time agents~\cite{wang2023voyager}—are bound by strict latency requirements to ensure fluid user interaction, but account for only a fraction of total compute demand. A massive and rapidly expanding volume of inference runs entirely \emph{offline} over pre-collected datasets, e.g., large-scale benchmark evaluation~\cite{hendrycks2020mmlu}, A/B testing of model variants~\cite{liang2022helm}, log and document processing~\cite{zheng2024batchllm}, retrieval index construction~\cite{karpukhin2020dense}, synthetic data generation for training~\cite{wang2022selfinstruct}, and reinforcement learning rollouts~\cite{ouyang2022gpt}. These jobs operate on pre-collected datasets with no human in the loop: individual requests may wait in a queue for seconds or minutes as long as the overall job meets a target wall-clock time and cost budget. As a result, the primary performance metric shifts from latency to aggregate \emph{throughput} (tokens per second), exposing new optimization opportunities for existing inference systems.

The throughput of an inference workload can be roughly characterized as $B / T$, where $B$ is the \emph{effective} batch size and $T$ is the per-\emph{iteration} decode time (offline jobs often spend most of their time in decoding~\cite{zheng2024batchllm}). Batching is therefore the primary lever for improving throughput: when $B$ is small, $T(B)$ typically grows sub-linearly as $B$ because larger batches amortize kernel launch overheads and memory accesses across more tokens, yielding higher arithmetic intensity and better overlap between compute and memory. As $B$ increases, the GPU eventually reaches a saturation point beyond which $T(B)$ grows roughly linearly with $B$; we denote this saturation batch size by $B_e$ (e.g., $B_e \approx 256$ for Qwen3-32B on H20 using tensor parallelism, see \S\ref{subsec:tradeoff_scalability_memory}). In offline workloads, the goal is therefore not to grow $B$ without bound, but to operate near this “efficient’’ regime around $B_e$, where the hardware is well utilized and further batching yields little marginal benefit.

However, pushing $B$ toward $B_e$ is tightly constrained by GPU memory capacity ($M$) because both static weights ($W$) and dynamic key-value (KV) state ($K$) must fit on each device. Under a fixed workload, the KV cache scales with the product of \emph{sequence length} ($S$) and batch size ($K = B S$)\footnote{We omit the constant per-token KV cache size in this equation and focus on the linear dependence on $B$ and $S$ for brevity.}. For modern multi-billion-parameter models, the weights alone already occupy a large fraction of $M$, so the remaining space for KV cache and other per-batch state quickly becomes the limiting factor for how far $B$ can grow.

This memory constraint poses a central challenge for distributed inference. Modern serving stacks typically combine data parallelism (DP) and model parallelism (MP) to fit larger models and scale throughput across nodes. DP offers excellent scalability for throughput workloads because replicas run largely independently, synchronizing only on infrequent control-plane signals (ideally $T' \approx \frac{T}{N}$ for $N$ GPUs); however, each DP replica fully replicates all model weights, leaving less memory for KV cache and thus limiting the achievable batch size ($B\approx N\frac{M - W}{S}$). MP shards weights across GPUs, reducing per-device parameter memory and enabling a larger batch size ($B \approx \frac{NM - W}{S}$), but introduces fine-grained, layer-wise communication and requires participating GPUs to march in lockstep (i.e., $T' \approx \alpha \frac{T}{N}$ with $\alpha > 1$). This tight coupling makes it difficult to exploit very large batches, and the gains from more aggressive MP can be offset by increased synchronization and reduced scheduling flexibility.
What is missing is a parallelism paradigm that combines the scalability and independence of DP (smaller $\alpha$) with the memory efficiency of sharded weights.

Viewed through the lens of resource economics, this dilemma naturally suggests an opportunity for \emph{resource arbitrage}: in inference deployments, high-speed interconnects often remain underutilized, while high-bandwidth GPU memory (HBM) capacity is extremely scarce. Can we deliberately trade a modest amount of otherwise idle bandwidth to “buy back’’ HBM from redundant weight replicas without sacrificing the scalability and simplicity of DP?

In this paper, we propose \sysname, Shared-weight Intra-node Data Parallelism tailored for offline LLM inference that reduces redundant weight replicas within a DP group. \sysname reorganizes the group into a distributed weight pool: instead of fully replicating all layers on every replica, each rank owns a disjoint subset of layers and exposes them to peers through one-sided, asynchronous access. Non-owner ranks fetch and cache these remote weights on demand, shrinking the per-GPU weight footprint and freeing HBM for KV cache and larger effective batches. 
To mitigate potential latency penalties when pulling weights in small-batch phases (whose prevalence we discuss in \S\ref{sec:limitations_fsdp}), \sysname introduces a dual-mode execution scheme based on this memory layout: a default \emph{Weight-as-a-Service} (WaS) mode that keeps computation local and streams weights, and a \emph{Compute-as-a-Service} (CaS) mode that instead ships activations to owners. The two modes are numerically equivalent but have different performance characteristics depending on arithmetic intensity.

We implement a prototype of \sysname on top of vLLM~\cite{kwon2023vllm} and evaluate it on several widely used GPU architectures, including H20, H200, and B200. Our experiments show that \sysname enables up to $1.8 \times$ larger KV capacity under the same memory budget and achieves $1.5 \times$ higher throughout compared to strong data- and tensor-parallel baselines.

\section{Background}
\label{sec:background}

In this section, we briefly review the architectural structure of modern LLMs, the execution model of offline inference, and the standard parallelism strategies used in practice to ground the design and analysis of \sysname in the rest of the paper.

\subsection{Large Language Models}

Modern large language models (LLMs) are predominantly based on decoder-only Transformers~\cite{vaswani2017transformer}. The network consists of an embedding layer, a stack of $L$ identical \emph{Transformer blocks}, and a final output projection. 

Each Transformer block (layer) is dominated by two submodules: (1) \emph{Self-attention}, typically implemented as multi-head self-attention with learnable query, key, value, and output projection matrices. (2) \emph{Feed-forward network (FFN)}, usually a two-layer MLP and a non-linearity in between.
The two dense linear layers in the FFN \emph{hold the majority of the weights} within each block, and, by extension, a large fraction of the total model parameters (e.g., 77\% in Qwen3-32B~\cite{yang2025qwen3} and 82\% in Llama-3.1-405B~\cite{dubey2024llama3.1}). 

Mixture-of-experts (MoE) models extend this structure by replacing the single FFN in each block with a set of expert MLPs and a router that selects a small subset of experts per token. In practice, however, offline workloads such as batch evaluation, log and document processing, synthetic data generation, and rollouts value stable behavior and still predominantly use dense LLMs~\cite{zhong2025streamrl,zhong2025rlhfuse,hu2024openrlhf} because MoE incurs additional routing, load-balancing, and communication complexity. Accordingly, this paper focuses on dense Transformer models; we discuss implications and potential extensions to MoE architectures in \S\ref{sec:conclusion}.

\subsection{Offline LLM Inference}
\label{subsec:autoregressive}
An offline inference job consists of a large, pre-defined set of requests (e.g., evaluation prompts, documents to summarize, or rollout trajectories) that must be processed by a pool of \emph{engines}. Each engine is typically bound to a fixed parallel configuration and runs LLMs in two phases: a \emph{prefill} phase that encodes the input context and populates the KV cache, and a \emph{decode} phase that generates tokens autoregressively. 
An inference \emph{engine}~\cite{zheng2024sglang, kwon2023vllm} maintains per-request state as a \emph{sequence} containing prompt tokens and generated tokens. Generation proceeds in discrete \emph{iterations}, each executing model \emph{forward} on GPUs and producing one new token per scheduled sequence. 
Within an engine, a \emph{scheduler} is responsible for \emph{token batching} and iteration-level decisions: it admits new requests, groups them into batches, interleaves prefill and decode work, and manages per-request KV cache and termination conditions. 
Above the engines, a \emph{job orchestrator} manages the job as a whole: it partitions the dataset into shards, assigns shards to engines, monitors progress and throughput, and may adjust global policies (e.g., target batch size and priority) based on runtime statistics. 

Offline workloads naturally expose opportunities for parallelism and batching, but also stress GPU memory and scheduling due to large KV caches and diverse request sizes. These characteristics differentiate offline from online inference, where requests are sparse, latency-sensitive, and batching opportunities are limited.

\subsection{Parallelism in LLM Inference}

To serve large models efficiently on multi-GPU systems, modern inference stacks employ a combination of data parallelism and model parallelism. We briefly summarize these strategies and the associated trade-offs.

\para{Data parallelism (DP).}
In data parallelism, each group of GPUs (i.e., an engine) holds a full copy of the model parameters. A batch of size $B$ is partitioned into $N$ shards for $N$ replicas, and each replica independently processes its local mini-batch of size $B/N$. During inference, replicas exchange only lightweight control-plane information (e.g., request routing, scheduling signals); the forward pass of each replica is otherwise independent. 

\para{Model parallelism (MP).}
Model parallelism~\cite{shoeybi2019megatron} partitions a single model across multiple GPUs so that one replica spans several devices, typically via tensor or pipeline parallelism. In tensor parallelism (TP), individual weight matrices are sharded across devices, each GPU computes on its shard, and collective operations assemble the final result, reducing per-device parameter memory at the cost of fine-grained, layer-wise synchronization. In pipeline parallelism (PP), the $L$ layers are split into contiguous stages assigned to different devices (or groups), and \emph{micro-batches} flow through these stages in a pipeline, trading reduced per-device memory for additional inter-stage communication and pipeline bubbles.

In practice, deployment stacks combine these techniques into a 2D or 3D parallelism configuration: (i) choose an MP layout (TP and, if necessary, PP) that leaves enough headroom for KV cache; then (ii) scale out with DP over such MP groups to increase throughput.
\section{Motivation}

\begin{figure}[t]
  \centering
  \includegraphics[width=\columnwidth]{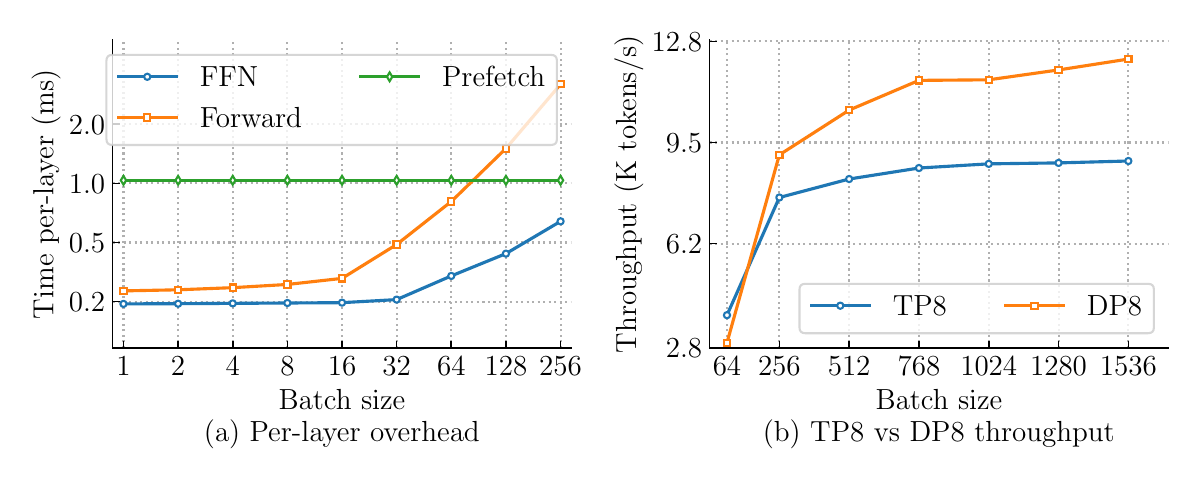}
  \caption{(a) Per-iter time (TP=2) with Llama-3.1-70B vs. $B$; (b) Throughput with Qwen3-32B vs. $B$. Both on H20, $S=$1K.}
  \label{fig:motivation1}
\end{figure}

\begin{figure}[t]
  \centering
  \includegraphics[width=\columnwidth]{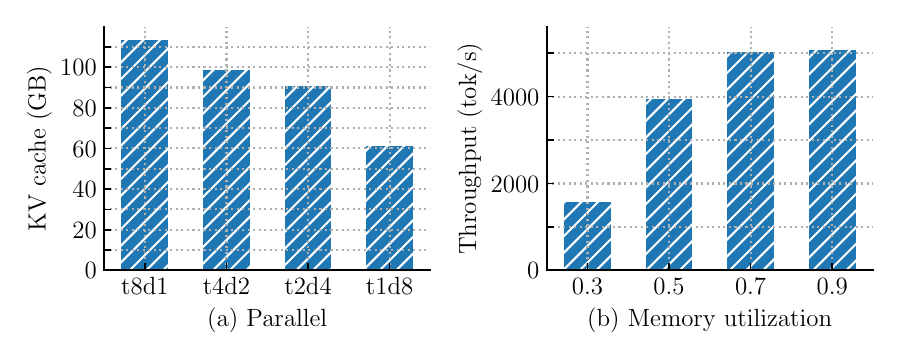}
  \caption{(a) KV cache size under different parallelism; (b) throughput under different memory utilization (TP=2, DP=4); on H20 with Qwen3-32B, $B=$1K, $S=$4K.}
  \label{fig:motivation3}
\end{figure}

To motivate \sysname, we first examine the main constraints of inference throughput, and then discuss why existing fully sharded data-parallel schemes fail to address these constraints in inference.

\subsection{Tradeoff in Scalability and Memory}
\label{subsec:tradeoff_scalability_memory}

We first discuss how batch size, compute scalability, and memory capacity jointly shape the inference throughput. 

\para{Sub-linear scaling of per-iteration time with batch size.}
In principle, the arithmetic work of Transformers scales linearly with $B$: each matrix multiplication and attention operation is applied to $B$ sequences at once. In practice, the measured per-iteration decode time $T(B)$ grows \emph{sub-linearly} with $B$ over a wide regime. As shown in Figure~\ref{fig:motivation1}(a), doubling $B$ from a small starting point increases $T(B)$ by less than $2\times$. This effect arises from several sources: kernel launch overheads and framework runtime costs are largely independent of $B$; GPU cores and tensor units are often underutilized at tiny batch sizes and become better saturated as $B$ grows; and memory accesses benefit from more efficient coalescing and higher arithmetic intensity when more tokens are processed together. As a result, the hardware moves from a latency-dominated regime to a more throughput-oriented regime as $B$ increases.

\para{Throughput improves with batch size until $B_e$.}
Because throughput is roughly $B / T(B)$, sub-linear growth of $T(B)$ in $B$ implies that $B / T(B)$ increases with $B$ over the same range. Figure~\ref{fig:motivation1}(b) illustrates this effect: going from very small batches to moderate batches yields large gains in throughput, while further increases yield progressively smaller marginal improvements as GPUs approach saturation. For example, on H20 with DP$=8$, increasing $B$ from $64$ to around $B_e\!\approx\!1{,}024$ improves throughput by $\sim4\times$, whereas pushing $B$ further from $1{,}024$ to $1{,}536$ only adds roughly $6\%$. Intuitively, higher DP degrees ($d$) require a larger $B_e$, because each replica only sees a local batch of size $B/d$, but across configurations the pattern is consistent: within a broad regime, increasing $B$ is the dominant knob for improving throughput on a fixed hardware budget.

\para{Throughput is constrained by available memory.}
The ability to increase batch size is limited by GPU memory capacity. For a fixed model and sequence length $S$, the KV cache footprint ($K$) grows linearly with $B S$ (as in \S\ref{sec:intro}). Empirically, for modern LLMs, $W$ already consumes a large fraction of $M$, so only a limited residual capacity is available for $K$; as $B$ grows and $K = B S$ approaches this residual capacity, the system enters a memory-pressure regime. Under data parallelism across $d$ replicas, the total replicated $W$ is $d$ times larger, so the system exhausts its residual KV capacity at a smaller $B$. Figure~\ref{fig:motivation3}(a) reveals this problem on an $8\times$ H20 testbed: as DP degree increases, the available KV cache size decreases by up to 45\%. That means even if $S=512$, this caps the effective batch at roughly $500$ sequences, which can still fall short of the saturation regime.
Many production engines avoid hard out-of-memory (OOM) failures by sequence preemption, throttling prefill, or refusing new requests, but these defenses still waste work and lower effective concurrency. Operators can try to stay away from this regime by capping $B$ or tightening admission control, yet this also leaves compute throughput on the table. In other words, whether we proactively shrink $B$ or let the runtime hover near the OOM cliff, effective throughput degrades when $K$ is insufficient. Figure~\ref{fig:motivation3}(b) illustrates this tension: for the same job, reducing GPU memory utilization from 0.9 to 0.3 erodes throughput by about 68\%.

These observations together illustrate a core tension in scaling LLM inference: from a compute perspective, we would like to keep increasing $B$ to move along the sub-linear region of $T(B)$ until $B_e$; from a memory perspective, large $W$ and $K$ tightly constrain how far $B$ can grow before capacity and management overheads negate these gains.

\subsection{Limited FSDP for inference}
\label{sec:limitations_fsdp}

\para{Fully sharded data parallel.} A natural way to reduce weight redundancy along the data-parallel dimension is to shard parameters across replicas, as in Fully Sharded Data Parallelism (FSDP)~\cite{zhao2023fsdp} and related ZeRO-style schemes~\cite{rajbhandari2020zerodp}. Conceptually, FSDP applies to inference much as it does to training: each rank stores only a shard of the parameters, and layers are materialized on demand via all-gathers before the forward pass. In high-throughput autoregressive serving, however, this mechanism faces two fundamental challenges.

\para{Challenges.}
The first concerns \emph{independence}. FSDP relies on group-wide collectives: all ranks participate in the same all-gather for a given layer at roughly the same time. However, inference engines are intentionally desynchronized due to continuous batching and variable sequence lengths. Enforcing step-synchronous all-gathers either stalls fast ranks or constrains the scheduler to move all replicas in lockstep, undermining the very flexibility that makes DP attractive.
The second challenge arises from the \emph{timescale and variability} of iterations. Training iterations last seconds with large, fairly stable mini-batches, so per-layer all-gathers can often be hidden under long forward–backward passes. Decode iterations, by contrast, are often only a few milliseconds and their effective batch size naturally shrinks toward the tail. In offline workloads, such small-batch regimes are common near the end of large evaluation or generation jobs, when only a small fraction of sequences remain active due to uneven sharding, filtering, or long-tail output lengths; empirical studies report that even 10\% of these long-tailed samples can account for up to half of total generation time~\cite{zhong2025rlhfuse}. In this regime, $T(B)$ collapses but FSDP still pays essentially the same all-gather cost to rebuild full weights on every rank, which can fail to hide behind computation. Thus, even without global synchronization, a “rebuild full weights on every rank at every iteration’’ pattern is too heavy-weight for millisecond-scale, dynamically shrinking batches.

These observations do not invalidate the core intuition of FSDP-style sharding—trading communication for memory—but substantially narrow the design space in inference. Any viable weight-sharing mechanism for high-throughput serving must (i) respect \emph{per-replica independence}, providing asynchronous, on-demand access to remote parameters, and (ii) adapt its communication pattern to the \emph{millisecond timescale} of decode iterations so that communication remains a small, overlap-friendly fraction of $T(B)$. 

\section{Design}
\label{sec:design}

\begin{figure*}[t]
    \centering
    \includegraphics[width=0.85\textwidth]{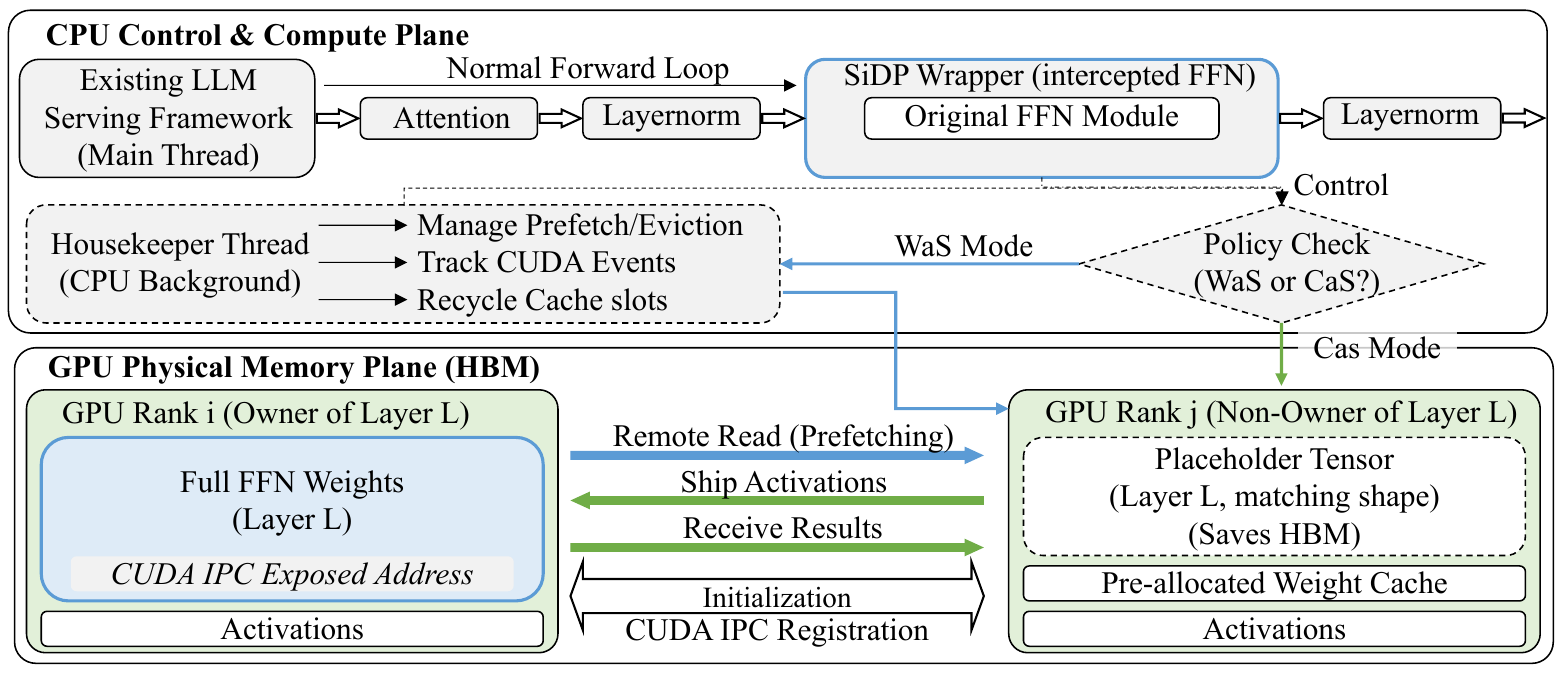}
    \caption{Overview of \sysname architecture and runtime workflow within a data-parallel groups.}
    \label{fig:arch}
\end{figure*}

The previous sections (\S\ref{subsec:tradeoff_scalability_memory} and \S\ref{sec:limitations_fsdp}) raise a central question: with bandwidth-backed resource in a DP group, how can we reduce redundant replicas to free HBM while preserving replica independence on the critical path of decoding iterations?

\sysname answers this question by decoupling the \emph{logical} semantics of data parallelism from the \emph{physical} placement of weights across GPUs. Each layer is assigned to an owner rank, and non-owner ranks access its FFN weights on demand through two complementary modes: WaS for the large-batch common case and CaS for small-batch tail phases.

\subsection{Overview}
\sysname is a thin wrapper around existing LLM serving frameworks (Figure~\ref{fig:arch}). It treats each Transformer FFN as a pluggable module: intercepting FFN forward passes, selecting WaS or CaS, and managing the corresponding weight/activation movement before invoking the original implementation. We focus on FFNs because attention weights are a much smaller parameter fraction, and remote attention is constrained by the locality and size of KV cache.
During initialization, \sysname parses the model to locate FFN layers, assigns an owner rank for each layer, and loads only the owner’s FFN weights on that GPU while allocating lightweight placeholders for non-owner layers. Owner ranks then register their FFN buffers and export the necessary handles so that peers can access the weights on demand, after which \sysname wraps the corresponding FFN forwards to enable transparent WaS/CaS execution at runtime. From the engine’s perspective, model architecture and numerics are unchanged; \sysname only changes where FFN weights reside and how they are accessed.



\begin{figure*}[t]
    \centering
    \includegraphics[width=\textwidth]{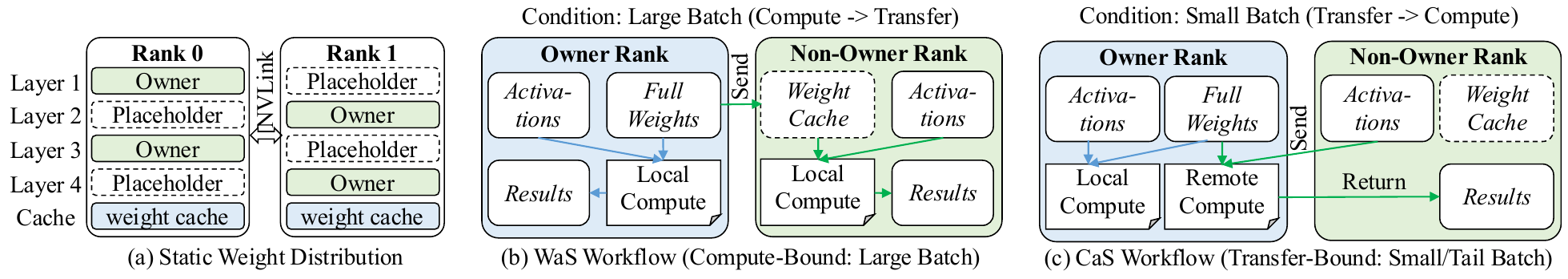}
    \caption{Memory layout and per-layer workflows of \sysname.}
    \label{fig:design}
\end{figure*}

\subsection{Weight as a Service}
\label{sec:was}

In the WaS mode, we exploit the fact that DP replicas on an NVLink-connected node share a high-bandwidth interconnect but do not actually need \emph{ownership} of the full model to execute a forward pass. The key observation, supported by Figure~\ref{fig:motivation1}(a), is that for a fixed layer the compute time dominates the cost of fetching that layer’s weights once the batch size is large enough. 
Instead of replicating a layer’s weights $d$ times across $d$ replicas, we can assign layer ($\ell$) to a single \emph{owner} rank $r(\ell)$  and allow the other replicas to fetch its parameters on demand over NVLink, while still performing the general matrix multiplications (GEMM) in FFN forward locally on their own activations, as shown in Figure~\ref{fig:design}. This preserves the independence and scheduling flexibility of data parallelism, while converting redundant weight replicas into a bandwidth-backed, shared resource inside the node.

\para{Asynchronous one-sided reading.}
To avoid the coordination and lockstep semantics of collective communication, WaS uses one-sided, asynchronous weight access. Initially, owner ranks register their weight buffers and expose device addresses that can be used by peer GPUs. When a non-owner rank needs layer $\ell$, it directly issues non-blocking device-to-device copies from $r(\ell)$’s HBM into its local cache. The owner is oblivious to the number and timing of these reads; it simply hosts the weights.

Because the compute for layer $\ell$ is performed locally, these remote reads can be issued ahead of time and overlapped with ongoing computation for other layers or tokens. In practice, the runtime maintains a small lookahead window and prefetches the next few layers while the GPU is executing forward for the current layer. This overlap hides NVLink latency and bandwidth fluctuations and avoids the ``all-or-nothing'' latency of reconstructing full layers via all-gathers. From the perspective of the serving engine, weight access becomes an asynchronous, per-rank operation that can be initiated whenever a given layer will soon be needed for a particular batch.

\para{Asynchronous buffer management.}
WaS relies on a small, fixed-capacity cache on each GPU to hold temporary remote weight. A naive implementation that repeatedly allocates and frees buffers per layer would quickly block the CPU control path. Instead, \sysname pre-allocates a set of cache slots and manages them with a dedicated housekeeper thread that runs asynchronously with respect to the main decode loop.
Each cache slot is associated with a particular set of \emph{tensor tiles} (e.g., MLP matrices of a layer). When the runtime decides to prefetch tiles for layer $\ell$, it reserves one slot, issues asynchronous remote reads into the slot, and attaches notifications so that the compute thread can wait on the completion of transfers just before launching computation. Once compute using that tiles has finished, the housekeeper observes the completion signals from the compute thread and marks the slot as free, making it available for the next prefetch wave. 
Because layers are visited in a predictable order, and the number of layers in flight is bounded by the computation progress, the cache can be dimensioned to keep only a small number of tiles per layer without risking thrashing.

\para{Peak shifting.}
When the data-parallel degree exceeds two, WaS introduces a new potential hotspot: for a given layer $\ell$, there is a single owner $r(\ell)$ that stores its weights, and all other DP ranks must fetch tiles from this owner. In the worst case, if $d$ replicas progress at similar speeds---a common situation in offline inference---then $d-1$ ranks may attempt to read the same weight tile from $r(\ell)$ at nearly the same time. This concentrates traffic on the owner's outgoing NVLink links and can reduce the effective bandwidth per reader to a small fraction of the theoretical peak (e.g., on DP $=8$, seven concurrent readers may each see less than $\tfrac{1}{7}$ of the link capacity).

To mitigate this, \sysname employs a decentralized, low-overhead \emph{peak shifting} scheme that staggers weight pulls across ranks without extra coordination. Conceptually, we provide each rank with enough cache to hold $(d-1)$ remote layers. We then organize layers by their indices into consecutive \emph{cycles} of size $d$. For a cycle that starts at layer $c$, rank $r$ applies a fixed per-rank offset anchored at this cycle start: it begins prefetching from layer $c + r$ and then proceeds in a wrap-around order within the same $d$-layer cycle (skipping its owner layer). 
This simple offsetting has two properties: (i) within each cycle, every rank can still obtain the full set of required non-owner tiles after at most $(d-1)$ prefetches, so it is never starved of weights; and (ii) at any given moment, different ranks tend to prefetch \emph{different} layers, avoiding many-to-one incast bursts. In effect, peak shifting distributes remote reads for each cycle over time and across ranks, preventing NVLink hotspots on owners while still ensuring that every replica fills up its cache with the remote weights it needs.

\subsection{Compute as a Service}
\label{sec:cas}
While WaS is well-suited to regimes where batch sizes are large enough to hide remote weight transfers, it becomes inefficient when the batch size $B$ is small (as in Figure~\ref{fig:motivation1}(a)). As discussed in \S~\ref{sec:limitations_fsdp}, such small-batch regimes arise naturally near the tail of jobs. 
In this phase, streaming full weight tiles to every non-owner rank can cost more bandwidth than is justified by the small amount of compute per rank.

CaS operates precisely in this tail regime by inverting the direction of data movement: instead of bringing weights to where activations reside, non-owner ranks send their activations to the owner of a layer, which performs the computation and returns the outputs, as shown in Figure~\ref{fig:design}. This introduces explicit point-to-point (P2P) interactions between replicas and therefore breaks the strict “everyone is fully independent” ideal of data parallelism. However, in the offline setting we target, CaS is only enabled when most replicas already have spare capacity—e.g., toward the end of a job when global concurrency is shrinking. In that regime, synchronizing a subset of ranks around shared compute is effectively a way of letting otherwise idle engines “help out” with the remaining work, rather than a restriction on progress in the throughput-critical bulk of the job.

\para{Skipping dummy runs.}
Mainstream inference engines~\cite{kwon2023vllm,zheng2024sglang} often keep all data-parallel replicas running dummy steps even when some of them have no active sequences. A common pattern is to run a dummy iteration with batch size $1$ on otherwise idle replicas so that internal schedulers, pipelines, and timing loops remain aligned across engines. In the tail of an offline job, this means that only a few replicas carry real work. Crucially, these dummy runs do not contribute any useful tokens and, in the context of CaS, should not trigger any real communication or compute.

\sysname makes this explicit by letting the job orchestrator declare whether engines are performing a dummy run.
For each layer, the owner only serves non-dummy work. When collecting activations from replicas, it ignores both P2P transfers and computation for those replicas whose state is dummy. Even if the owner carries dummy tasks, it still executes the fused computation for received activations and returns outputs to the corresponding ranks. 
Non-owner ranks adopt the dual behavior. If their work is dummy, they immediately return from the layer without packing activations or initiating any communication; the dummy step is satisfied purely at the control level, with no data movement. 
This design allows existing engines to retain their dummy-run rhythm for liveness, while ensuring that CaS pays no additional bandwidth or compute cost for those dummy iterations.

\para{GEMM fusion.}
The effectiveness of CaS hinges on whether the owner can serve many non-owner requests without suffering from small-batch inefficiencies. Our motivation measurements (Figure~\ref{fig:motivation1}(a)) show a crucial property: the FFN forward time at small batch sizes grows very slowly with $B$ and quickly saturates once a modest batch size is reached. In other words, adding more rows has negligible marginal cost over a wide range until the GEMM reaches a certain utilization level. CaS exploits this by fusing activations from multiple ranks into larger owner-side GEMMs.

Concretely, CaS maintains a staging buffer per layer $\ell$ into which it concatenates incoming activation chunks from all non-owner ranks (and its local activations). Once all activations are ready, the owner launches a single GEMM and then partitions the output along the same boundaries and returns each slice to its originating rank. P2P transfers are issued by non-owners and can be overlapped with the computation of earlier steps in the decode iteration. Thus, CaS leverages both batch fusion at the owner and overlap between P2P transfers and upstream computation to make “compute sharing” efficient in the small-batch tail of offline jobs.

\para{Consistent mode switching.}
Since CaS introduces cross-rank interactions, it must be enabled in a controlled, globally consistent way. In \sysname, the job orchestrator decides when to switch between WaS and CaS. The orchestrator monitors progress and load across all DP replicas—tracking, for example, how many sequences remain active per engine and the observed batch sizes in recent iterations. When it detects that a job has entered a tail regime where many engines are underutilized, it can broadcast a control-plane directive for all engines to operate in CaS mode.
This centralized policy ensures that all ranks agree on the communication pattern for each layer and avoids inconsistent mode choices that could lead to deadlocks. Mode switches occur only at coarse-grained boundaries (empirically minute-level at the tail of a job) with hysteresis to avoid flapping so the high-throughput bulk of the job runs purely in WaS and preserves full DP independence. 

In practice, the orchestrator bases this decision on a hardware-specific batch-size threshold $B_{\text{th}}$ derived from offline profiling. For each GPU and model type, we estimate the minimum batch size at which forward time can fully hide the cost of weight fetching in WaS, given the devices’ compute and NVLink bandwidth (as validated in \S\ref{sec:evaluation}). As long as the effective batch size stays above $B_{\text{th}}$, WaS remains preferable; when it persistently falls below $B_{\text{th}}$, the orchestrator switches the group into CaS mode.

\subsection{Discussion}
\para{Implementation.} \sysname is realized as a plugin for \texttt{vLLM}~\cite{kwon2023vllm} without requiring any modifications to the original vLLM source code. The implementation is lightweight, comprising roughly 4,000 lines of Python, and integrates seamlessly as a drop-in extension. \sysname uses CUDA IPC to expose local GPU memory for remote access. All weight/activation copies happen on separate CUDA streams to avoid blocking main computation. 
Notifications between housekeepers and main compute threads are driven by CUDA events for explicit dependencies without synchronous CPU barriers.
In practice, \sysname maintains $d-1$ slots for WaS mode and two slots for CaS mode, costing $\le 1$ GB GPU memory for mainstream LLMs. 

\para{Deployment scope.}
\sysname is designed as an intra-node primitive that operates within a DP group on NVLink-connected GPUs: ownership, WaS/CaS decisions, and weight sharing are all confined to a single node’s parallel group. However, this does not preclude multi-node deployments: large-scale serving stacks already structure clusters as collections of identical 2D/3D-parallel groups and scale out by replicating these groups across machines without cross-group communication. In such settings, \sysname applies unchanged inside each node-local group and multi-node scaling is simply achieved by replicating \sysname-enabled groups at the job orchestrator level. 

\para{Robustness.} From a robustness perspective, \sysname is closer to model parallelism than conventional DP: within a \sysname-enabled group, FFN weights are owned by specific ranks, so the failure domain resembles that of TP/PP where all participating GPUs are required for correctness. Consequently, replacing a TP degree with an equivalent \sysname degree does not materially change robustness, while applying \sysname to “convert” DP capacity can weaken the fault-isolation benefits that DP would otherwise provide. This trade-off remains local to each group and can be mitigated at cluster scale by increasing the number of replicated \sysname-enabled groups.
\section{Evaluation}
\label{sec:evaluation}

This section answers four questions: (i) How does KV cache capacity change under our design? (ii) Does \sysname improve end-to-end throughput across models and testbeds? (iii) How effective is the design of \sysname in practice? (iv) When does CaS outperform WaS, and how sensitive is overall performance to the WaS/CaS mode switch?

\begin{figure}[t]
  \centering
  \includegraphics[width=\columnwidth]{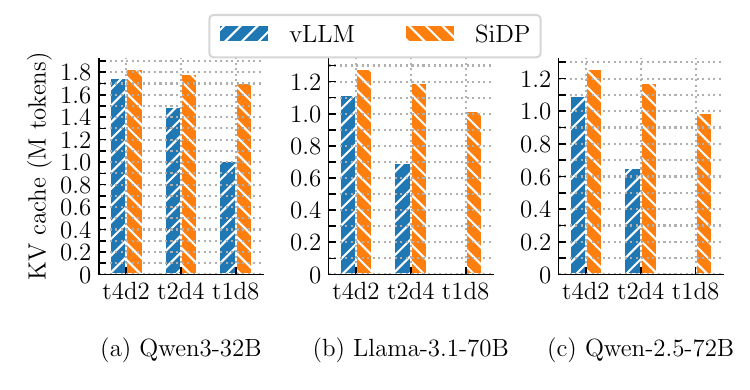}
  \caption{Available KV cache size (in tokens) of {\tt vLLM} vs. \sysname under different parallelism on H20. {\tt vLLM} cannot run Llama-3.1-70B and Qwen2.5-72B with TP=1, DP=8.}
  \label{fig:kv}
\end{figure}

\begin{figure*}[t]
  \centering
  \includegraphics[width=\textwidth]{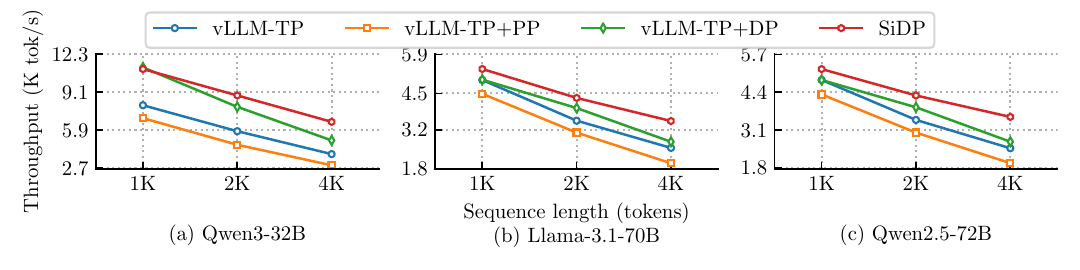}
  \caption{End-to-end throughput with different models, parallelisms, and sequence lengths on H20.}
  \label{fig:e2e-h20}
\end{figure*}

\begin{figure*}[t]
  \centering
  \includegraphics[width=\textwidth]{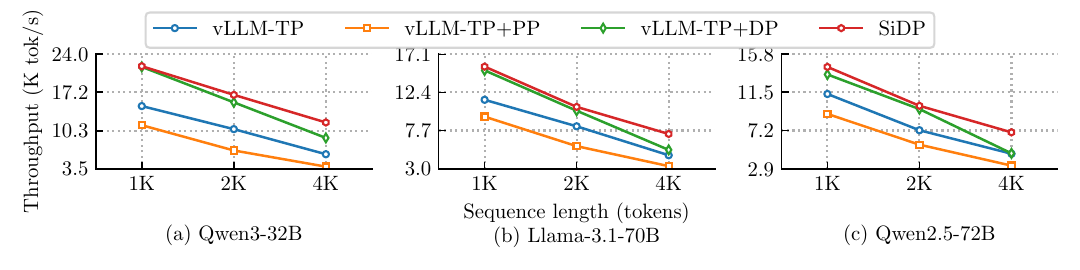}
  \caption{End-to-end throughput with different models, parallelisms, and sequence lengths on H200.}
  \label{fig:e2e-h200}
\end{figure*}

\begin{figure*}[t]
  \centering
  \includegraphics[width=\textwidth]{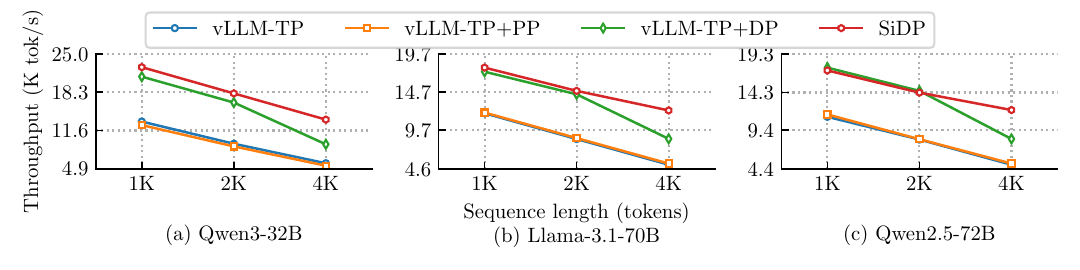}
  \caption{End-to-end throughput with different models, parallelisms, and sequence lengths on B200.}
  \label{fig:e2e-b200}
\end{figure*}

\subsection{Evaluation Setup}
\begin{table}[t]
\caption{Evaluation testbed overview.}
\label{tab:testbed}
\centering
\renewcommand{\arraystretch}{1.12}
\setlength{\tabcolsep}{3pt}
\begin{tabularx}{\columnwidth}{|>{\raggedright\arraybackslash}p{0.26\columnwidth}|X|X|X|}
\hline
\textbf{Config} & \textbf{H20 node} & \textbf{H200 node} & \textbf{B200 node} \\
\hline
GPU model & $8 \times$ H20 & $8 \times$ H200 & $8 \times$ B200 \\
\hline
GPU memory & 144 GB & 144 GB & 180 GB \\
\hline
Interconnect & NVLink~4 & NVLink~4 & NVLink~5 \\
\hline
CUDA version & 12.6 & 12.6 & 12.8 \\
\hline
\end{tabularx}
\end{table}

\para{Testbed.} Our evaluation is conducted on the following testbeds: H20, H200, and B200, as shown in Table~\ref{tab:testbed}.

\para{Baseline and models.} We compare \sysname against {\tt vLLM} (0.10.1)~\cite{kwon2023vllm}. We choose three popular models served in offline inference: Qwen3-32B~\cite{yang2025qwen3}, Qwen2.5-72B~\cite{qwen2.5}, and Llama-3.1-70B~\cite{dubey2024llama3.1}.

\para{Parallelism.} We use three types of parallelism configurations: pure tensor parallelism, hybrid TP+DP, and hybrid TP+PP and report the best performance in each category. Since \sysname reduces weight redundancy in data parallelism and supports DP8 for all evaluated models, we show its results at DP=8 by default. 

\para{Workload.} We select three typical offline inference workloads: summarization~\cite{see2017get} (average $S=$1K), code generation~\cite{chen2021evaluating} ($S=$2K), and conversation evaluation~\cite{kwan2024mteval} ($S=$4K). For each workload, we try two batching strategies: (1) a fixed large batch size to fully utilize GPU compute but may trigger OOM protection of engines; (2) an adaptive batch size ($B \approx \frac{M - W}{S}$). We report the better of the two settings. 

\subsection{End-to-end Performance}
\label{sec:e2e}

\para{KV cache capacity.}
Figure~\ref{fig:kv} quantifies the KV cache size each configuration can sustain (in tokens) under a fixed memory budget, and directly tests the memory model from \S\ref{subsec:tradeoff_scalability_memory}: reducing the per-device weight footprint $W$ should enlarge the residual KV cache capacity. Across all three models and DP/TP settings, \sysname systematically increases the maximum KV token budget compared to {\tt vLLM} (${\sim}1.7\times$ with TP=2, DP=4.  Even more strikingly, for the largest DP setting (TP=1, DP=8), {\tt vLLM} cannot sustain Llama-3.1-70B or Qwen2.5-72B at all under our memory limit, while \sysname continues to support ${\approx}1.0$M KV tokens.
The gains are modest at low DP or smaller models (e.g., about $5\%$ for Qwen3-32B at TP=4, DP=2) where redundant FFN replicas are not the dominant HBM consumer. As DP increases and weight replication becomes the main consumer of memory, the benefits grow rapidly, explaining the much larger gains at TP=2,DP=4 and TP=1,DP=8. These trends match the expectation that, as DP grows and weight replication becomes more dominant, sharding FFN weights along the DP axis frees a growing fraction of HBM for KV cache. The expanded KV capacity directly enables larger feasible batch sizes and/or longer supported sequences under the same hardware and model configuration.

\para{Throughput.}
Figure~\ref{fig:e2e-h20},~\ref{fig:e2e-h200}, and~\ref{fig:e2e-b200} show how \sysname converts its KV-capacity advantage into higher throughput across GPUs, models, and workload lengths. Compared to a TP-only baseline, adding DP in {\tt vLLM} improves tokens/s by roughly $1.0{\sim}1.8\times$; we also include a TP+PP baseline for completeness, which is generally below TP in our single-node setting. \sysname further boosts the DP+TP configuration by expanding the residual KV capacity and enabling larger effective batches. The effect is most pronounced in the long-context, memory-stressed regime ($S{=}4\text{K}$): across H20/H200/B200, \sysname achieves roughly $1.3$--$1.5\times$ higher throughput than DP+TP.
The benefit of \sysname grows with both model size and sequence length: at short sequences (1K), where the workload is already largely compute-bound, the DP baseline can sustain aggressive batching without hitting KV limits and \sysname closely matches its throughput. As we move to 2K and especially 4K workloads, the DP baseline enters a memory-bound regime and can no longer increase batch size without triggering OOM protection or preemption, whereas \sysname can continue to scale $B$ and stay in the sub-linear region of $T(B)$. The end result is that, for memory-bound offline workloads, the best-performing configuration with \sysname lies strictly above any TP/PP/DP combination available in the {\tt vLLM} baseline under the same hardware and model. 

The consistent gains across three GPU generations suggest this advantage is not tied to a specific device. Looking ahead, even though future GPUs will likely offer larger HBM, they will also bring higher compute capability and better kernel efficiency, which can push the saturation batch size $B_e$ upward. In this light, \sysname should remain a useful way to expand the feasible batch-size region and preserve headroom for higher-throughput operating points.

\subsection{Design Choice Validation}

\begin{figure}[t]
  \centering
  \includegraphics[width=0.9\columnwidth]{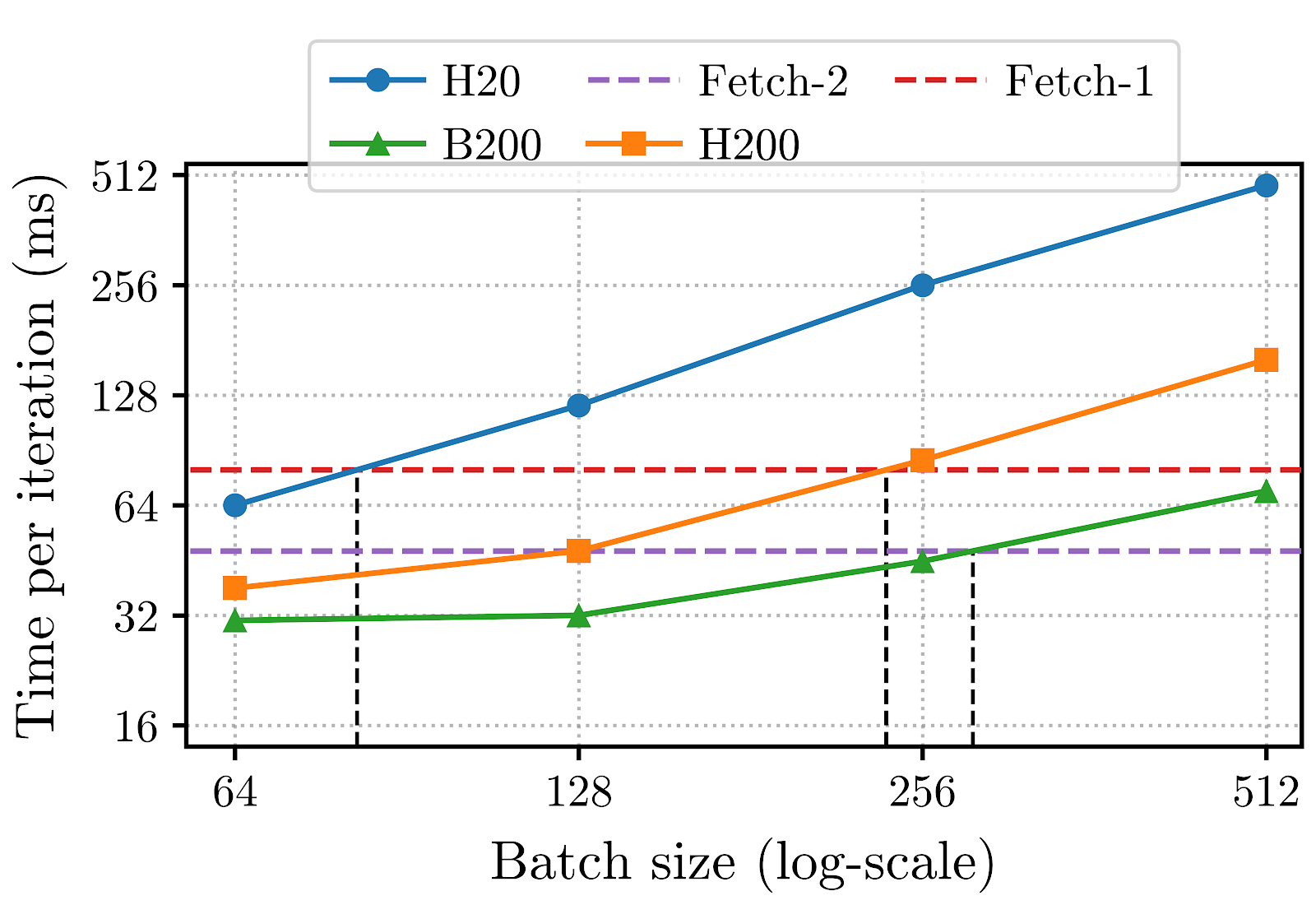}
  \caption{Prefetch vs. $T(B)$ (TP=2, $S=$ 1K). Fetch-1 and 2 are the time to read entire FFN weights of Llama-3.1-70B on NVLink~4 and NVLink~5. }
  \label{fig:prefetch}
\end{figure}

\para{Prefetch is overlappable.}
Figure~\ref{fig:prefetch} tests a key assumption behind WaS: on modern GPUs, the time to fetch FFN weights can be overlapped with decode computation. For each batch size, we report the per-iteration decode time $T(B)$ on {\tt H20}, {\tt H200}, and {\tt B200}, alongside a pessimistic estimate of the time needed for \sysname to fetch \emph{all} FFN weights across all layers (``Fetch-1''/``Fetch-2'' for NVLink~4/5). As $B$ grows from 64 to 512, $T(B)$ increases while the fetch time remains essentially constant. On H20, $T(B)$ already exceeds the full-weight fetch time at $B{=}128$ and is roughly $3{\sim}5\times$ larger at $B{=}256$ and $512$. On H200, $T(B)$ slightly surpasses the fetch time at $B{=}256$ and reaches about $1.8\times$ at $B{=}512$. Even on the faster B200, $T(B)$ approaches or exceeds the B200 fetch time near $B=256$. These trends corroborate our expectations from \S\ref{subsec:tradeoff_scalability_memory}: in the large-batch regime where WaS is intended to operate, decode iterations become compute-dominated, while the cost of weight prefetch stays bounded and can be hidden within $T(B)$ on all three GPU generations.

The experiment is intentionally conservative: the ``Fetch'' lines reflect streaming \emph{all} FFN weights once, whereas the runtime only fetches non-owner layers (a $(d{-}1)/d$ fraction). The fact that $T(B)$ is already comparable to or much larger than this worst-case full-fetch time indicates ample slack to overlap prefetch, even when NVLink bandwidth is shared with other traffic.

\begin{figure}[t]
  \centering
  \includegraphics[width=\columnwidth]{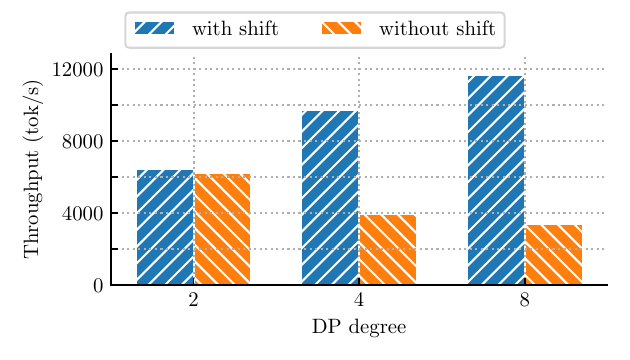}
      \caption{Impact of peak shifting with Qwen3-32B on H20, $S=$1K.}
  \label{fig:peak}
\end{figure}

\begin{figure}[t]
  \centering
  \includegraphics[width=\columnwidth]{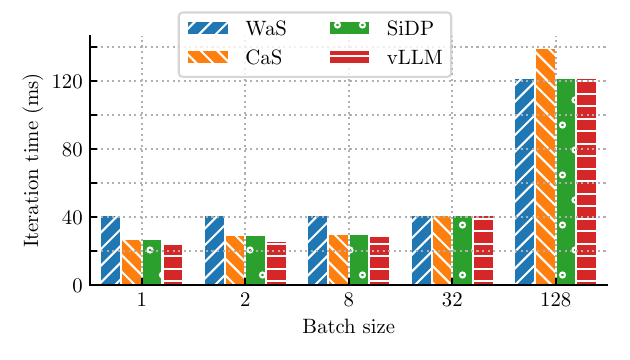}
  \caption{Iteration time of WaS/CaS/\sysname/{\tt vLLM} with Llama-3.1-70B on H20, TP=2, DP=2.}
  \label{fig:mode}
\end{figure}

\para{Impact of peak shifting.}
Figure~\ref{fig:peak} evaluates how peak shifting affects end-to-end throughput as we increase the DP degree under \sysname. This experiment directly targets the NVLink incast problem discussed in \S\ref{sec:was}: without any staggering, all non-owner ranks tend to pull the same layer tiles from the same owner at roughly the same time, concentrating traffic on a single set of links. With peak shifting, we permute the order in which ranks prefetch tiles so that, at any instant, different ranks are typically reading tiles from \emph{different} owners. The results show that this decentralized staggering is crucial once DP exceeds a small group size. At DP=4, peak shifting boosts throughput by roughly $2.5\times$, and at DP=8 the improvement grows to $3.4\times$. This confirms that avoiding incast on owners is necessary to translate WaS’s HBM savings into DP-scalable throughput on NVLink-connected nodes.

\para{Mode switching mitigates tail inefficiency.}
Figure~\ref{fig:mode} reports per-iteration decode time under different batch sizes for WaS-only, CaS-only, the {\tt vLLM} baseline, and \sysname (which effectively takes $\min\{\text{WaS},\text{CaS}\}$ at each $B$). This experiment directly tests the design goal in \S\ref{sec:cas}: WaS should dominate in the large-batch interior of an offline job, while CaS should mitigate the small-batch tail where WaS can no longer hide weight transfers. The data shows a clear crossover. For small batches ($B \le 32$), CaS is substantially faster than WaS, reflecting that moving activations to a shared owner is more efficient than streaming full weights to every rank when per-rank work is tiny. For large batches ($B{=}128$), WaS matches the {\tt vLLM} baseline, while CaS pays extra P2P and fusion overhead. 

For users, the key takeaway is that a simple mode-switching policy suffices to capture the best of both worlds, as we discussed in \S\ref{sec:cas}. There remain small gaps between \sysname and the {\tt vLLM} baseline at the tiniest batch sizes ($B{=}1$): $12\%$ slower due to the fixed control overhead of routing activations through the CaS wrapper and coordinating owner-side fusion. These differences are confined to the extreme tail, where the number of active sequences—and thus the number of iterations spent in this regime—is small relative to the bulk of the job.

\begin{figure}[t]
  \centering
  \includegraphics[width=\columnwidth]{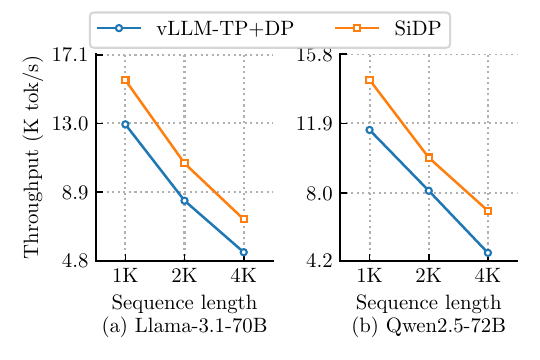}
  \caption{End-to-end throughput on H200 with GPU memory utilization $=0.6$.}
  \label{fig:short}
\end{figure}

\para{\sysname works for short-context scenarios.}
A natural concern is that \sysname might only help in long-context settings. To stress-test this, we run Llama-3.1-70B and Qwen2.5-72B with the same configuration as \S\ref{sec:e2e} but under a tighter memory budget: $0.6$ GPU memory utilization (about $80$\,GB per GPU), roughly matching common A100/H100 deployments, as in Figure~\ref{fig:short}. For each model and sequence length, we sweep DP/TP configurations and report the best throughput for the {\tt vLLM} DP+TP baseline and for \sysname (we omit pure TP and TP+PP schemes for brevity). \sysname improves throughput by $24\sim 51\%$ across all evaluated sequence lengths.
These results show that \sysname is not confined to extreme long-context workloads. Once the model is large enough that weights consume a substantial fraction of HBM under realistic budgets, reducing the effective weight footprint along the DP axis still enlarges the feasible batch-size region and translates into higher throughput—even when sequence lengths are only one or two thousand tokens.

\subsection{Ablation}

\begin{figure}[t]
  \centering
  \includegraphics[width=\columnwidth]{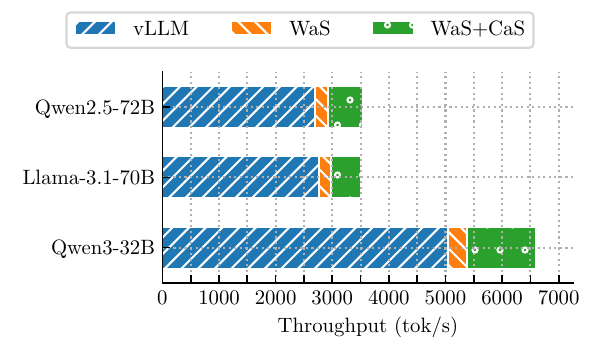}
  \caption{Throughput gain of \sysname with/without mode switch on H20 when running different models, DP=8.}
  \label{fig:switch}
\end{figure}

\para{Impact of mode switch.}
Figure~\ref{fig:switch} quantifies how much of \sysname's end-to-end gain comes from WaS alone versus the combined WaS+CaS design on H20 ($S$=4K). With WaS only, \sysname improves over {\tt vLLM} by about $7\%{\sim}9\%$, confirming that the overhead of weight transfer in small-batch cases can still erode overall performance. When we enable mode switching, the gains become substantially larger ($27\%{\sim}32\%$ over {\tt vLLM}).

These results show that WaS and CaS play complementary roles at the job level. Adding CaS does not change the steady-state batch size in the interior—where WaS remains active—but it significantly reduces the cost of the tail phase, where effective batch size collapses and WaS can no longer hide weight transfers. Importantly, we do not observe any configuration where enabling CaS reduces overall throughput relative to WaS-only on this 4K workload, indicating that the orchestrator’s coarse-grained, tail-triggered switching policy is effective: CaS is invoked where it helps (small-batch regimes) and stays out of the way where WaS already matches or beats the baseline. This confirms that mode switching is not just a micro-optimization for per-iteration latency, but a meaningful contributor to end-to-end job throughput.

\begin{figure}[t]
  \centering
  \includegraphics[width=\columnwidth]{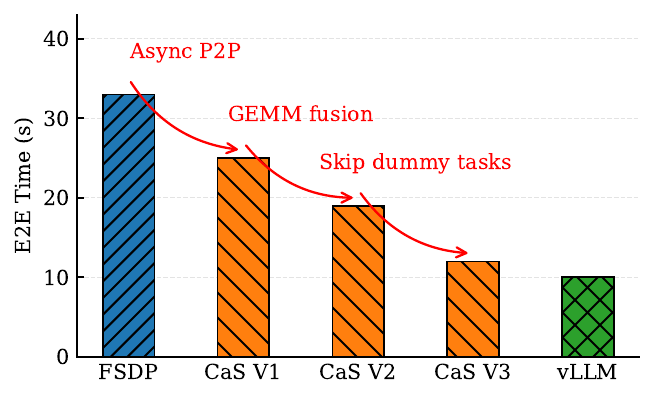}
  \caption{End-to-end time of a single request under different implementations with Llama-3.1-70B on H20, TP=2, DP=2.}
  \label{fig:cas}
\end{figure}

\begin{figure}[t]
  \centering
  \includegraphics[width=\columnwidth]{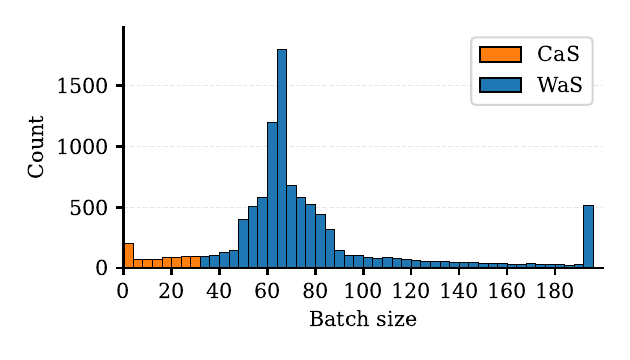}
  \caption{Batch/mode distribution of \sysname with Llama-3.1-70B on H20, DP=8, $S{=}4\text{K}$.}
  \label{fig:hist}
\end{figure}

\para{Impact of CaS optimizations.}
Figure~\ref{fig:cas} breaks down the impact of successive CaS optimizations in the most challenging regime for weight sharing: batch size $B{=}1$ in the long tail of an offline job. A naive FSDP-style scheme that all-gathers full FFN weights for each layer on every rank (``FSDP'') takes 33\,s to finish the tail workload, over $3\times$ slower than the {\tt vLLM} baseline. This gap illustrates the limitation of FSDP in inference discussed in \S\ref{sec:limitations_fsdp}. Starting from this baseline, we progressively refine CaS while still \emph{using shared weights}.

The first variant (``CaS V1'') replaces collectives with asynchronous P2P, reducing tail time from 33\,s to 25\,s ($24\%$ faster). ``CaS V2'' adds fused GEMMs on the owner, concatenating activations from multiple ranks into a single large GEMM and cutting time further to 19\,s (another $24\%$ reduction). Finally, ``CaS V3'' incorporates dummy skipping (\S\ref{sec:cas}), avoiding any P2P or GEMM work for replicas that only execute dummy iterations and reducing the tail from 19\,s to 12\,s (${\sim}37\%$). 

Taken together, these optimizations improve a straightforward FSDP-like implementation by $2.8\times$. This experiment shows that \sysname is not only conceptually better suited to inference than FSDP-style sharding, but that careful implementation choices—asynchronous P2P, GEMM fusion, and dummy skipping—are all necessary to make weight sharing practical in the most challenging corner of offline serving.

\para{Tail characteristics.} 
We finally address a potential concern: does \sysname only accelerate offline jobs for a short period while leaving the long tail largely unoptimized? 
The answer is no. 
We profile a representative long-context job ($S{=}4\text{K}$) and record the effective batch size and selected mode at each iteration in Figure~\ref{fig:hist}. 
Using the hardware-specific threshold $B_{\text{th}}$ derived in \S\ref{sec:cas}, we find that more than $90\%$ of iterations remain in the WaS-enabled regime. 
This clarifies an important distinction: \emph{a long tail does not imply a long CaS phase}. 
Instead, CaS acts as a short but essential safety net for the tail-of-the-tail. 
Although this window is brief, keeping its performance close to {\tt vLLM} is critical because small-batch iterations can contribute disproportionately to overall job completion time. 
This also explains why the combined WaS+CaS design outperforms WaS-only in \S\ref{sec:evaluation}, even though CaS is exercised for only a small fraction of the execution.

\section{Related Work}

\para{Token batching.}
Token batching and scheduling are central to exploiting GPU parallelism in autoregressive LLM inference. Orca~\cite{yu2022orca} introduces \emph{continuous batching} and iteration-level scheduling so that batches can evolve across decode iterations, substantially improving utilization under variable-length generation. Follow-up systems~\cite{2024Sarathi,DeepSpeed,ge2024fastgen,he2024fineinfer} refine these ideas with phase-aware batching, speculative or hierarchical schedulers, and cost-aware ordering to better handle mixed or bursty workloads. \sysname is orthogonal to these techniques: it focuses on weight memory, enabling larger per-GPU batches that token-batching schedulers such as Orca, vLLM, and their successors can exploit.

\para{Weight memory management.}
Another line of work attacks weight memory pressure via slower memory tiers or cluster-level pooling. ZeRO-Infinity~\cite{ren2021zero} and FlexGen~\cite{shen2023flexgen} offload weights and activations to CPU or NVMe and carefully prefetch to compensate for limited device memory. More recently, Pie~\cite{xu2024pie} pools CPU memory for inference, while KunServe~\cite{cheng2024kunserve} treats parameters as a cluster-level resource and elastically rebalances them across GPUs. Oneiros~\cite{li2025oneiros} instead optimizes KV cache for multi-tenant serving by remapping parameters so that co-located models share KV space more efficiently. These systems primarily rely on CPU or storage bandwidth and sophisticated planning to make offloaded weights usable at scale. \sysname instead assumes a high-bandwidth interconnect and redistributes weights \emph{within} the GPU group, converting redundant replicas into extra on-device capacity for KV/cache without paying PCIe/NVMe latency.

\para{Offline serving optimization.}
Recent work has begun to treat offline, throughput-oriented serving as a first-class workload. Glinthawk~\cite{hamadanian2025glinthawk} and TD-Pipe~\cite{zhang2025td} decouple phases of the model (e.g., attention vs. MLP, or different pipeline stages) to better pack compute and KV/cache across heterogeneous resources, while BatchLLM~\cite{zheng2024batchllm} and BlendServe~\cite{zhao2024blendserve} use global prefix trees, resource-aware batching, and latency-relaxed reordering to improve utilization. SamuLLM~\cite{fang2025samullm} uses a sampling-then-simulation planner to schedule multi-LLM offline applications.
These systems mostly optimize request ordering, prefix reuse, and KV/cache placement on top of a fixed parallelism substrate. \sysname is complementary: it changes the weight-memory layout inside each DP group, enlarging the feasible batch-size regime that such offline schedulers can then leverage.

\section{Conclusion}
\label{sec:conclusion}
Offline LLM inference is dominated by throughput (tokens per second), but effective batch size is jointly constrained by compute saturation and memory: larger batches move the system toward a more efficient regime, yet KV cache $K \propto BS$ and fully replicated weights $W$ under data parallelism quickly exhaust HBM, while model parallelism reduces per-device $W$ at the cost of DP-style scalability. \sysname addresses this tension by treating FFN weights as a bandwidth-backed shared resource inside each DP group, organizing them as a distributed pool where each layer is owned by a single GPU and other replicas access its parameters on demand via a Weight-as-a-Service mode for large batches and a Compute-as-a-Service mode for small-batch tails. On NVIDIA H20, H200, and B200, this design increases usable KV capacity by up to $1.8\times$ under the same DP/TP configurations and translates that into up to $1.5\times$ higher end-to-end throughput.

\para{Limitations and future work.}
This work focuses on dense, decoder-only Transformers deployed on NVLink-connected systems. Extending \sysname to heterogeneous clusters and slower interconnects (e.g., PCIe or Ethernet) will require revisiting the balance between WaS and CaS, adapting prefetch policies to more constrained bandwidth, and co-designing with cross-node scheduling. Another promising direction is support for MoE architectures. Modern MoE systems often rely on expert-parallel load balancing (EPLB) to exploit redundancy and equalize load across experts; \sysname could complement these techniques by using WaS-style weight sharing to reduce redundant expert replicas within an expert-parallel group. We leave the design of such an \sysname-aware EPLB layer as promising future work.

\bibliographystyle{plain}
\bibliography{reference}

\appendix

\end{document}